\documentclass[aps,preprint]{revtex4} 
\usepackage{amsmath}
\usepackage{graphicx}

\newcommand{\etal}{et al. }
\newcommand{\jpb}{J. Phys. B: At. Mol. Opt. Phys. }
\newcommand{\jpa}{J. Phys A: Math. Theor. }
\newcommand{\pr}{Phys. Rev. }

\begin{document}

\title{Asymmetric steering in coherent transport of atomic population with a three-well Bose-Hubbard model}

\author{M.K. Olsen}\email{Corresponding author: mko@physics.uq.edu.au}
\affiliation{School of Mathematics and Physics, University of Queensland, Brisbane 4072, Australia}

\begin{abstract}

We analyse the Einstein-Podolsky-Rosen (EPR) steering properties in a three-well Bose-Hubbard model under the mechanism of coherent transport of atomic population (CTAP). This process, also known as transport without transit (TWT), transfers condensed atoms from one well of a lattice potential to another, non-adjacent well, without macroscopic occupation of the middle well. The dynamics and entanglement properties of this system have previously been analysed and shown to be dependent on the initial quantum state of the atoms in the first well. In this work we utilise the phase-space method of stochastic integration in the positive-P representation. The system demonstrated inseparability between the wells for Fock initial states, and we show that steering is also only found for an initial Fock state. Depending on how we write our steering inequalities, steering is seen at different times, showing a basic asymmetry. This means that for part of the evolution, Alice can steer Bob but not vice versa. After the midpoint of the population transfer, this situation reverses.
  
\end{abstract}


\maketitle 

\section{Introduction}
\label{subsec:intro}

The possibility of macroscopic matter transport without transit (TWT) of a Bose-Einstein condensate (BEC) was first raised by Rab \etal~\cite{Rab2008,Bradly2012} and Opatrn\'{y} and Das~\cite{Opatrny2009}. This phenomenon is derived from the process of Stimulated Raman Adiabatic Passage (STIRAP), originally proposed to transfer atomic population between two atomic levels of a three-level atom without macroscopically populating an intermediate level~\cite{Oreg1984,Kuklinski1989,Bergmann1998}. This is performed via the application of a counter-intuitive sequence of electromagnetic pulses to the atoms. Considering an atomic system with levels denoted by $|1\rangle,\: |2\rangle $ and $|3\rangle$, where the object is to transfer population from $|1\rangle$ to $|3\rangle$, a pulse denoted by $K_{23}$ dresses the $|2\rangle$ to $|3\rangle$ transition before a pulse denoted by $K_{12}$ is applied to the transition between $|1\rangle$ and $|2\rangle$. The result when everything works as proposed is that the population is transferred adiabatically from the first to the third state without ever populating the second.

Since the original proposal for the transfer of population between atomic levels, STIRAP has also been proposed for the conversion between atomic and molecular BEC, where there is one atomic state linked to an excited molecular state and finally a stable molecular state~\cite{Mackie2000,Hope2001,Drummond2002,Mackie2004,Ling2004}. A related Raman system with trapped BEC was proposed and analysed for a squeezed atom laser~\cite{Haine2005a,Haine2005b}, the production of atom-light entanglement~\cite{Haine2006}, quantum state measurement of an atom laser~\cite{Bradley2007}, and statistics swapping and quantum state transfer~\cite{Olsen2008}.

In this paper we will analyse the process of TWT in a three-well Bose-Hubbard system~\cite{Nemoto2000}, consisting of three potential wells in a linear configuration, where the condensed atoms are initially all in one of the end wells and at the final time have been transferred to the well at the other end, without significantly populating the middle well~\cite{Rab2008}. This system has previously been analysed using a Bose-Hubbard model~\cite{Rab2008,Bradly2012}, and a Gross-Pitaevskii model~\cite{Rab2008} which includes the spatial extent of the three wells with interactions between the atoms, as well as a Schr\"odinger equation approach with spatial extent but without interactions~\cite{Opatrny2009}. 
What we add here, by using a phase-space representation approach, is the quantum noise due to both the initial quantum statistics and the interactions, which allows us to calculate any deviations from the mean-field predictions~\cite{Chianca4well,Chiancathermal}, as well as the quantum correlations between the modes.

This work extends an earlier analysis of the system~\cite{CTAPmuzza} which used the positive-P~\cite{P+} and truncated Wigner representations~\cite{Graham1973} to analyse both the mean-field dynamics and entanglement properties of this system in terms of initial quantum states. It was found that whether the initial population was treated as a coherent or Fock state made no difference to the mean-field dynamics, but the results for the entanglement inequality used were different, with no entanglement being discovered for initial coherent states. In this work we will present results in a regime where the positive-P representation gives stable results. 

We wish to detect the presence of EPR (Einstein-Podolsky-Rosen)-steering, first predicted by Einstein \etal~\cite{EPR} with what became known as the EPR paradox, and discussed by Schr\"odinger~\cite{Erwin1,Erwin2} using a slightly different approach. In 1989 Reid formulated the EPR paradox in mathematical terms using the quadrature amplitudes of the electromagnetic field, which have similar properties to the position and momentum originally considered~\cite{Margaret}. This was demonstrated experimentally shortly afterwards by Ou \etal~\cite{Ou}. More recently, Wiseman \etal showed that there was an equivalence between the phenomena described by Einstein and Schr\"odinger, and put them on a firm mathematical foundation~\cite{Wisesteer}. Wiseman \etal raised the question of whether asymmetric steering was possible, given that having one system steer another has an implicit asymmetry. In the case of a three-mode system, Olsen \etal had already answered this affirmatively in the case of Gaussian measurements~\cite{tripart}. Midgley \etal~\cite{sapatona} and Olsen~\cite{meu} have theoretically analysed optical systems where asymmetric Gaussian steering is predicted theoretically, while Handchen \etal~\cite{Natexp} demonstrated this phenomenon experimentally. In this work we link the three-well atomic system with steering. Using different inequalities~\cite{HZ,ericsteer} which do not depend on any Gaussian assumptions, we show that asymmetric bipartite steering is also possible, with the asymmetry being apparent in time.   

\section{System and Hamiltonian}
\label{sec:sysHam}

\begin{figure}[tbhp]
\centerline{\includegraphics[width=0.8\columnwidth]{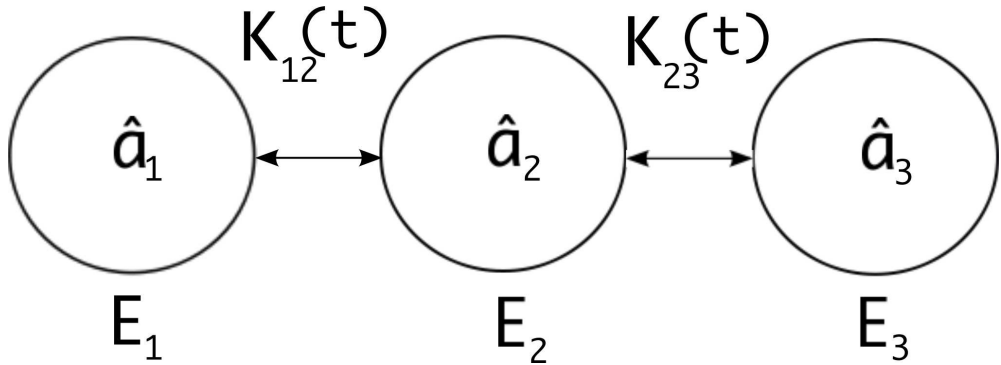}}
\caption{Simpified schematic of the three well system. The $\hat{a}_{j}$ are bosonic annihilation operators for atoms in the $j$th well, the $E_{j}$ are the single-atom energy levels of each well, and the $K_{ij}(t)$ are the time dependent tunnelling strengths between wells $i$ and $j$.}
\label{fig:scheme}
\end{figure}

Before we detail how we will analyse steering, we will give a description of the physical system, since this will influence the inequalities that we will use.
We may extend the procedure followed by Milburn \etal~\cite{Milburn1997} to treat our three-well system~\cite{Nemoto2000} as containing a single mode per well. A schematic is shown in Fig.~\ref{fig:scheme}, for which we can write the Hamiltonian,
\begin{eqnarray}
{\cal H}/\hbar &=& \sum_{j=1}^{3}\left( E_{j}\hat{a}_{j}^{\dag}\hat{a}_{j} + \chi \hat{a}_{j}^{\dag\;2}\hat{a}_{j}^{2} \right)
- K_{12}(t)\left(\hat{a}_{1}^{\dag}\hat{a}_{2}+\hat{a}_{1}\hat{a}_{2}^{\dag}\right) \nonumber \\
& & - K_{23}(t)\left(\hat{a}_{2}^{\dag}\hat{a}_{3}+\hat{a}_{2}\hat{a}_{3}^{\dag}\right),
\label{eq:Ham}
\end{eqnarray}
where the $\hat{a}_{j}$ are the bosonic annihilation operators for atoms in each of the modes, the $E_{j}$ are the ground-state energies of each well, $\chi$ represents the collisional interactions, and the $K_{ij}(t)$ represent the time dependent couplings between the wells. Following Rab \etal~\cite{Rab2008}, we set $E_{1}=E_{3}=0$, $E_{2}>0$, with $K_{12}(t) = \Omega\sin^{2}\left[\pi t/2t_{p}\right]$ and $K_{23}(t) = \Omega\cos^{2}\left[\pi t/2t_{p}\right]$, where $t$ runs from $0$ to the pulse time, $t_{p}$. These parameters, with $\Omega = 10$, $t_{p}=400/\Omega$,  $E_{2}=\Omega$, and $-.005\leq \chi\leq 0.005$, were found to give good adiabaticity of population transfer, with very little population to be found in mode $2$ at any one time. We used a total number of atoms of $N_{A}=N_{1}+N_{2}+N_{3}=200$. We note here that our Hamiltonian, although written in a slightly different way, is equivalent to that of Rab \etal~\cite{Rab2008}.

From the Hamiltonian our preferred option is to use the positive-P representation~\cite{P+}, which allows for the exact calculation of normally ordered expectation values for systems described by this type of Hamiltonian, but can suffer from severe stability problems when the interaction term becomes dominant~\cite{Steel,unstableP+}. In our previous analysis of CTAP we also used mean-field and truncated Wigner representation equations, but this is not necessary here because we are only analysing a regime where we know the positive-P equations to be stable. We will use the flexibility inherent in the phase space representation to model the initial state of the condensate in the first well as both a coherent state and a Fock state of fixed number~\cite{DFW}, in order to calculate the effects of the initial quantum statistics on both the subsequent dynamics and the quantum correlations of interest.

\section{Analysis of EPR steering}
\label{sec:EPR}

In the present system we are interested in entanglement between the atomic modes at each site, rather than between individual atoms. While this seems obvious because the atoms are indistinguishable bosons and only the spatially separated modes can be distinguished, we state it here to avoid any possible confusion. In particular, we wish to investigate the possibility of steering between the modes in wells $1$ and $3$, since these are the two in which the atomic populations are found. In practice this means that we are interested in whether the density matrix of the combination of modes can be written as a combination of density matrices for the individual modes\cite{CVCpure}, which follows the original idea as expounded by 
Schr\"odinger. For continuous-variable systems, the canonical methods use the system covariance matrix to develop inequalities based on the partial positive transpose criterion~\cite{Duan, Simon}, but we have previously found these to not be entirely suitable for systems with a $\chi^{(3)}$ nonlinearity, due to rotation of the Wigner function in phase-space~\cite{nonGauss}, which means that the quadrature angle for best measurement changes dynamically. While measurement phase can be optimised, we feel it is better to use a measure that is independent of phase.

In our previous analysis we used the work of Hillery and Zubairy~\cite{HZ}, which presents a wider range of inequalities, the violation of which can be used to demonstrate two-mode entanglement. Hillery and Zubairy used a Cauchy-Schwarz inequality to show that (using our notation),
\begin{equation}
\langle N_{1}N_{3}\rangle < |\langle \hat{a}_{1}\hat{a}_{3}^{\dag}\rangle |^{2},
\label{eq:HandZ}
\end{equation}
means that there exists bipartite entanglement between the atomic modes in wells $1$ and $3$. Hilary and Zubairy show that this is a stronger measure than the Duan-Simon inequalities which are so useful for bipartite Gaussian entanglement, detecting entanglement in cases where it is missed by those measures. 
The results of Hillery and Zubairy were subsequently extended by Cavalcanti \etal~\cite{ericsteer}  to show that, with $j$ numbering the modes, the level of correlation required to demonstrate EPR-steering in an N-mode system is given by
\begin{equation}
|\langle\prod_{j=1}^{N}\hat{a}_{j}\rangle |^{2} > \langle N_{1}\prod_{j=2}^{N}(N_{j}+1/2)\rangle.
\label{eq:Eric}
\end{equation}
For the two modes of interest and since where the numbering begins from is arbitrary this allows us to define the functions
\begin{eqnarray}
\xi_{13} &=& \langle\hat{a}_{1}^{\dag}\hat{a}_{3}\rangle\langle\hat{a}_{3}^{\dag}\hat{a}_{1}\rangle - \langle \hat{a}_{1}^{\dag}\hat{a}_{1}\left(\hat{a}_{3}^{\dag}\hat{a}_{3}+1/2\right)\rangle, \nonumber \\
\xi_{31} &=& \langle\hat{a}_{1}^{\dag}\hat{a}_{3}\rangle\langle\hat{a}_{3}^{\dag}\hat{a}_{1}\rangle - \langle \hat{a}_{3}^{\dag}\hat{a}_{3}\left(\hat{a}_{1}^{\dag}\hat{a}_{1}+1/2\right)\rangle,
\label{eq:2steer}
\end{eqnarray}
which are very straightforward to calculate in the positive-P representation as they are already in normal order. We see immediately that the choice of numbering means that there is an inherent asymmetry here, as with the Reid variance criteria. If Alice does the measurements on mode $1$ and Bob does them on mode $3$, $\xi_{13}>0$ signifies that Bob would be able to steer Alice, but not the other way around. $\xi_{31}>0$ shows that the nonlocality is such that Alice can steer Bob, but again not vice versa. Therefore for symmetric steering both functions need to be greater than zero at the same time. We note here that the expectation values in the above functions can in principle be measured, using the techniques developed by Ferris \etal~\cite{Andyhomo} for any phase-dependent terms.

\section{Phase-space equations of motion and quantum state simulation}
\label{sec:EoM}

In order to develop the equations of motion for the positive-P phase-space variables, we proceed via the standard techniques~\cite{Crispin}, mapping the Hamiltonian onto a master equation, then the appropriate Fokker-Planck equations. Following the standard correspondence rules, we find the coupled stochastic differential equations in the positive-P representation, which necessitates six variables in a doubled phase-space to maintain positivity of the Fokker-Planck equation diffusion matrix,
\begin{eqnarray}
\frac{d\alpha_{1}}{dt} &=& -2i\chi\alpha_{1}^{2}\alpha_{1}^{+} + i\Omega\sin^{2}\left(\pi t/2t_{p}\right)\alpha_{2}+\sqrt{-2i\chi\alpha_{1}^{2}}\;\eta_{1}(t), \nonumber \\
\frac{d\alpha_{1}^{+}}{dt} &=& 2i\chi\alpha_{1}^{+\;2}\alpha_{1} - i\Omega\sin^{2}\left(\pi t/2t_{p}\right)\alpha_{2}^{+} + \sqrt{2i\chi\alpha_{1}^{+\;2}}\;\eta_{2}(t), \nonumber \\
\frac{d\alpha_{2}}{dt} &=& -i\left(E_{2}+2\chi\alpha_{2}^{+}\alpha_{2}\right)\alpha_{2} + i\Omega\left[\sin^{2}\left(\pi t/2t_{p}\right)\alpha_{1}\right. \nonumber \\
& &\left. +\cos^{2}\left(\pi t/2t_{p}\right)\alpha_{3}\right] +\sqrt{-2i\chi\alpha_{2}^{2}}\;\eta_{3}(t), \nonumber \\
\frac{d\alpha_{2}^{+}}{dt} &=& i\left(E_{2}+2\chi\alpha_{2}^{+}\alpha_{2}\right)\alpha_{2}^{+} - i\Omega\left[\sin^{2}\left(\pi t/2t_{p}\right)\alpha_{1}^{+}\right. \nonumber\\
& &\left.+\cos^{2}\left(\pi t/2t_{p}\right)\alpha_{3}^{+}\right] +\sqrt{2i\chi\alpha_{2}^{+\;2}}\;\eta_{4}(t), \nonumber\\
\frac{d\alpha_{3}}{dt} &=& -2i\chi\alpha_{3}^{2}\alpha_{3}^{+} + i\Omega\cos^{2}\left(\pi t/2t_{p}\right)\alpha_{2} + \sqrt{-2i\chi\alpha_{3}^{2}}\;\eta_{5}(t), \nonumber \\
\frac{d\alpha_{3}^{+}}{dt} &=& 2i\chi\alpha_{3}^{+\;2}\alpha_{3} - i\Omega\cos^{2}\left(\pi t/2t_{p}\right)\alpha_{2}^{+} + \sqrt{2i\chi\alpha_{3}^{+\;2}}\;\eta_{6}(t).
\label{eq:P+SDE}
\end{eqnarray}
In the above, the $\eta_{j}$ are normal Gaussian noise terms, such that $\overline{\eta_{j}(t)}=0$ and $\overline{\eta_{j}(t)\eta_{k}(t')}=\delta_{jk}\delta(t-t').$ The averaging of the solutions of these equations over many stochastic trajectories approach normally-ordered operator expectation values, with, for example
\begin{equation}
\overline{\alpha_{j}^{+\;m}\alpha_{k}^{n}} \rightarrow \langle \hat{a}_{j}^{\dag\;m}\hat{a}_{k}^{n}\rangle ,
\label{eq:averages}
\end{equation}
whenever the integration is stable and converges. The positive-P representation is well known for instability and divergence properties, especially with $\chi^{(3)}$ nonlinearities~\cite{Steel,unstableP+}, but we will only be presenting results here where it has reliably converged.

The two initial quantum states which we will use are the Glauber-Sudarshan coherent state, which is the closest quantum state to a classical state of fixed amplitude and phase, and the Fock state, which has a fixed number but totally indeterminate phase~\cite{states}. We have chosen these two because, if TWT is phase dependent, we would expect them to lead to the most different results. Both have been found previously to lead to mean-field results for three-well Bose-Hubbard systems which only match the classical predictions for short times~\cite{Chianca4well,Chiancathermal} and we found that the inseparability properties were also state dependent~\cite{CTAPmuzza}. A coherent state, $|\alpha\rangle$, with coherent excitation $\alpha$, which includes the vacuum state, $|0\rangle$, is particularly easy to model.  In the positive-P representation, one of the ways to sample $|\alpha\rangle$ is
\begin{equation}
\alpha_{P} = \alpha, \:\:\: \alpha_{P}^{+}=\alpha^{\ast}.
\label{eq:alphaP}
\end{equation}

A Fock state is a little more difficult. In the positive-P representation a Fock state can be sampled from
\begin{eqnarray}
\alpha_{P} &=& \mu+\gamma, \nonumber \\
\alpha_{P}^{+} &=& \mu^{\ast}-\gamma^{\ast},
\label{eq:FockPplus}
\end{eqnarray}
where $\gamma=(\eta_{1}+i\eta_{2})/\sqrt{2}$, with the $\eta_{j}$ being normal Gaussian variables, and is easily sampled using the standard methods. The variable $\mu$ is found from a Gamma distribution for $z=|\mu|^{2}$,
\begin{equation}
\Gamma(z,N+1)=\frac{\mbox{e}^{-z}z^{N}}{N!},
\label{eq:Gammadist}
\end{equation}
using a method given by Marsaglia and Tsang~\cite{Marsaglia}, then taking $\mu=\sqrt{z}\;\mbox{e}^{i\theta}$, where $\theta$ is uniform on $[0,2\pi)$.

\section{Quantum dynamics}
\label{sec:dynamics}

In order to benchmark our results against what is already available, we have used parameters very similar to those of Rab \etal~\cite{Rab2008}. Our average number of atoms is $200$, with these all being found in the first well at $t=0$, and we have used $\Omega = 10$, with $t_{p}=400/\Omega$, and $E_{2}=0.1\Omega$ and $\chi=10^{-4}$ in all our calculations. Setting $E_{2}>0$ means that the coupling rate out of well $2$ into the adjacent wells is higher than that into it, thus helping to keep the well population as low as possible. We change the quantum state of the atomic mode in the first well, finding no discernible effect on the mean-field dynamics.

\begin{figure}[tbhp]
\begin{center} 
\includegraphics[width=0.9\columnwidth]{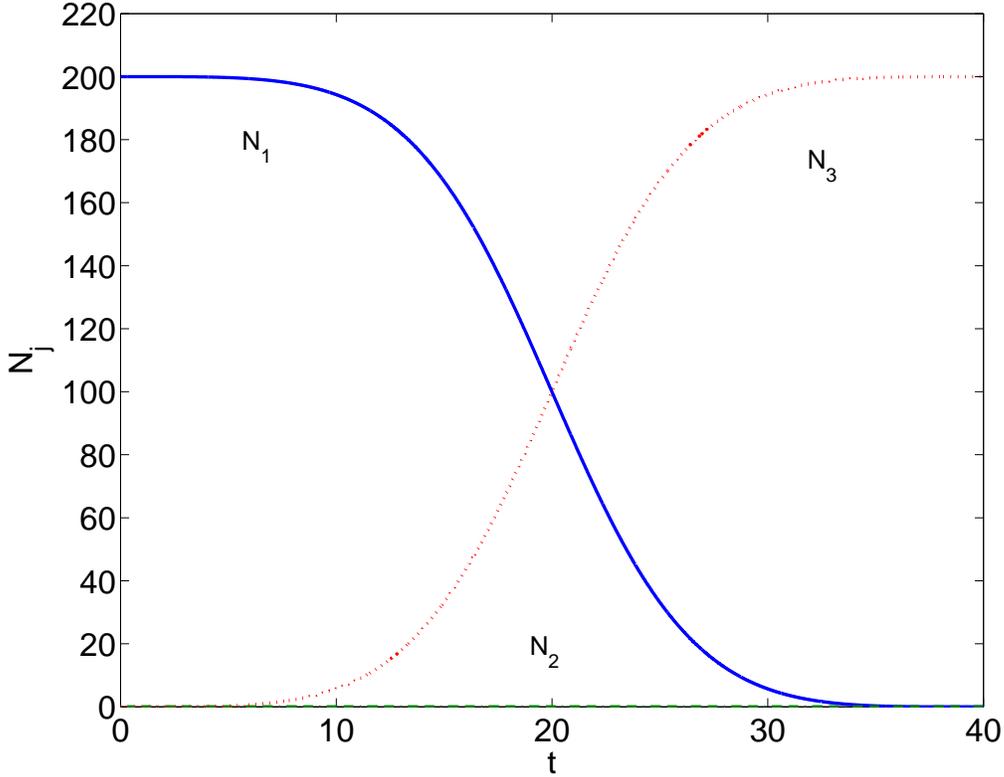}
\end{center} 
\caption{The populations in each of the three wells as a function of time (in units of $10/\Omega$). This graph was calculated using $3.53\times 10^{5}$ trajectories, with $\Omega=10$, $E_{2}=1$, $\chi = 10^{-4}$, $N_{1}(0)=200$, $t_{p}=40$, and $N_{2}(0)=N_{3}(0)=0$. $N_{1}$ is initially in a coherent state. For these parameters, an initial Fock state in the first well gives an indistinguishable result.}
\label{fig:populations}
\end{figure}

We next examine the steering properties of the system, concentrating on steering between the modes in the first and third wells. The middle well is not of interest here, as we require its population to stay as low as possible. It is obvious from inspection of Eq.~\ref{eq:2steer} that the initial values of $\xi_{13}$ and $\xi_{31}$ will be different, 
with $\xi_{13}(t=0)=-N_{1}(0)/2$ and $\xi_{31}(t=0)=0.$ As with ordinary entanglement, we therefore expect to find any steering to be present during the middle period of the evolution, when the two outside wells are both populated. We note here that, due to the presence of the $\chi^{(3)}$ collisional nonlinearity and the lack of Gaussian assumptions in the derivation of the $\xi_{ij}$, any steering is non-Gaussian. If steering followed the same sort of development in time as the entanglement examined in Ref.~\cite{CTAPmuzza}, we might expect it to be maximised half way through the coherent population transfer process, when the atoms are equally distributed between the two end wells. Assuming unrealistically for a moment that the quantum states remain the same throughout the evolution, we can find analytic solutions for the $\xi_{ij}$ at this half way point. For both coherent and Fock states we find that the inequalities are not violated, but the functions are markedly negative. For coherent states, we find $\xi_{ij}=-N_{T}/4$ and for Fock states, we find $\xi_{ij}=-N_{T}/4(N_{T}+1)$, where $N_{T}=N_{1}(0)$ is the total number of atoms in the system. This non-violation is, of course, totally expected since any entanglement means that the initial quantum states have to change.

\begin{figure}[tbhp]
\begin{center} 
\includegraphics[width=0.9\columnwidth]{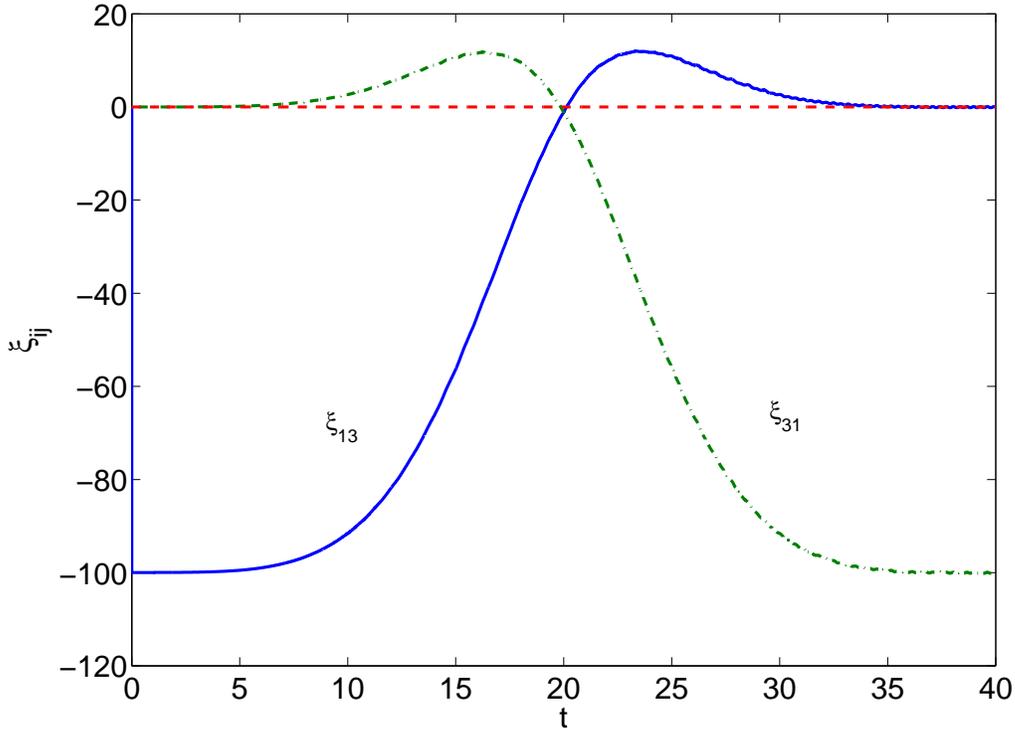}
\end{center} 
\caption{The EPR-steering correlation, $\xi_{13}$ of Eq.~\ref{eq:2steer}, calculated for an initial Fock state in mode $1$. The solid line is $\xi_{13}$ and the dash dotted line is $\xi_{31}$. This result is for $5.5\times 10^{4}$ trajectories. Parameters are as for Fig.~\ref{fig:populations}. The dashed line at $0$ is a guide to the eye. We see that Bob can steer Alice for a period of time, but this changes to Alice being able to steer Bob after the half way point.}
\label{fig:PPxi}
\end{figure}

\begin{figure}[tbhp]
\begin{center} 
\includegraphics[width=0.9\columnwidth]{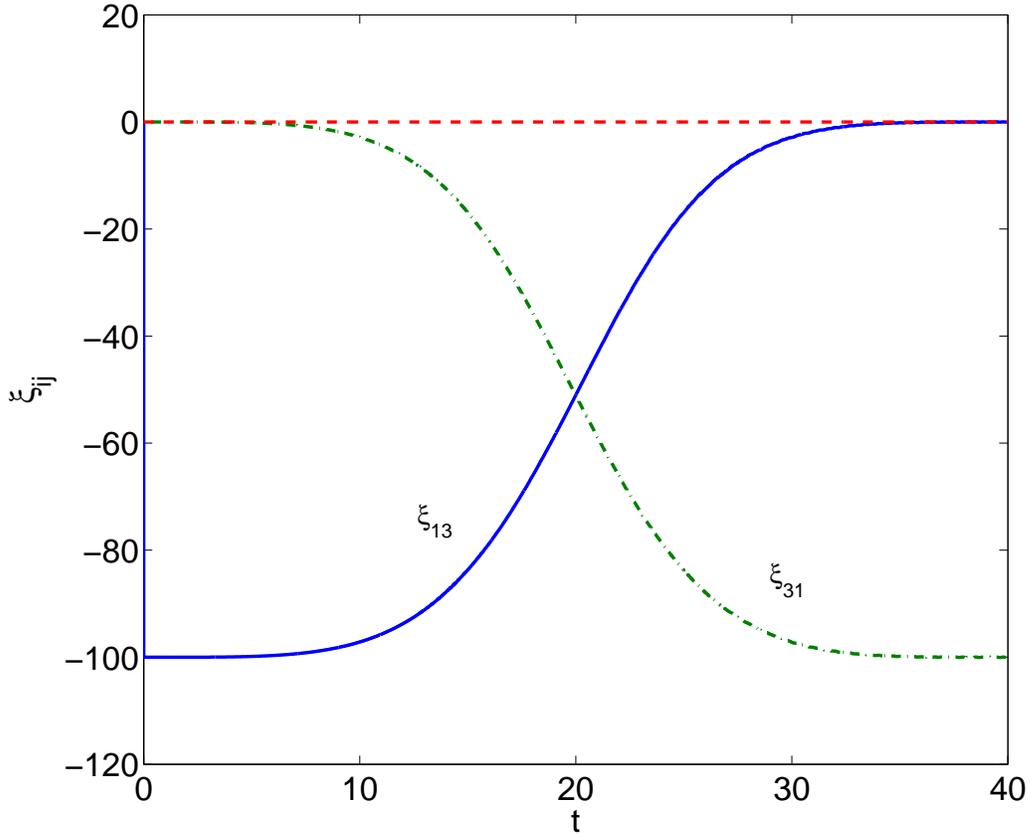}
\end{center} 
\caption{The EPR-steering correlation, $\xi_{13}$ of Eq.~\ref{eq:2steer}, calculated for an initial coherent state in mode $1$. The solid line is $\xi_{13}$ and the dash dotted line is $\xi_{31}$. This result is for $5.5\times 10^{4}$ trajectories. Parameters are as for Fig.~\ref{fig:populations}. The dashed line at $0$ is a guide to the eye.}
\label{fig:PPcoherent}
\end{figure}

Fig.~\ref{fig:PPxi} shows the results of the steering inequalities for an initial Fock state in the first well, while Fig.~\ref{fig:PPcoherent} shows them for an initial coherent state, all for the same parameters as used in Fig.~\ref{fig:populations}. We see that steering is predicted for the initial Fock states, but at different times for the two functions, depending on how we label the modes. This asymmetry therefore is evident in time, with no instance where both the $\xi_{ij}$ are greater than zero. For the initial coherent state, both functions are always negative and we see no evidence of the degree of nonlocality needed for steering. This is expected, since we found no entanglement at all for this situation in our earlier work~\cite{CTAPmuzza}, and steering is a subset of entanglement.

With the variance inequalities developed by Reid~\cite{Margaret}, it is obvious from the derivation whether mode $i$ is steering mode $j$ or vice versa. This is also the case here, as outlined by Jones \etal~\cite{steerJones}. With these functions developed from the inequalities of Cavalcanti \etal~\cite{Cavalcanti}, the mode labelled $j$ in $\xi_{ij}$ is the one on which measurements can potentially steer the other. 
Therefore we have shown that the system exhibits the degree of non locality necessary to exhibit steering in an asymmetric manner, with the mode which can steer the other at any given time during the population transfer process swapping over at the half way point. 

\section{Conclusions and discussion}
\label{sec:conclude}

In conclusion, we have performed fully quantum analyses of the process of transport without transit for a three-well Bose-Hubbard type model where all the population is initially in one well. We find that the initial quantum statistics, although not having a noticeable effect on the mean-field behaviour, do become important when we consider the nonlocal properties of the two end wells. The dynamics following from an initial coherent state remain separable and hence we cannot see the degree of nonlocality needed for EPR-steering. However, an initial Fock state leads to entanglement of the atomic modes in the two end wells and also exhibits EPR-steering. There is a temporal asymmetry 
evident in this steering, with it becoming apparent at different times depending on the labelling of the modes. This signifies that there is an asymmetry in terms of which of the modes can steer the other at various times.
 As well as presenting interesting quantum dynamics, this simplified system is a source of non-Gaussian entangled states which are physically separated, composed of massive constituents, and possess a sufficient degree of nonlocality for steering. This could well make it useful for fundamental investigations of quantum information science with massive particles, with its usefulness depending on how closely a real system could approximate the three-mode model.

\section*{Acknowledgments}

This research was supported by the Australian Research Council under the Future Fellowships Program (Grant FT100100515).

\end{document}